\documentclass[twocolumn,showpacs,preprintnumbers]{revtex4}
\usepackage{amssymb}
\usepackage{amsfonts}
\usepackage{amsmath}
\usepackage{graphicx}
\usepackage{dcolumn}
\usepackage{bm}
\usepackage{epsfig}

\begin{document}

\preprint{APS/123-QED}
\title{Kinetic description of rotating Tokamak plasmas \\
with anisotropic temperatures in the collisionless regime}
\author{Claudio Cremaschini}
\altaffiliation[Also at ]{Consortium for Magnetofluid Dynamics, University of Trieste, Italy}
\affiliation{International School for Advanced Studies (SISSA) and INFN, Trieste, Italy}
\author{Massimo Tessarotto}
\altaffiliation[Also at ]{Consortium for Magnetofluid Dynamics, University of Trieste, Italy}
\affiliation{Department of Mathematics and Informatics, University of Trieste, Italy}
\date{\today }

\begin{abstract}
A largely unsolved theoretical issue in controlled fusion research is the
consistent \textit{kinetic} treatment of slowly-time varying plasma states
occurring in collisionless and magnetized axisymmetric plasmas. The
phenomenology may include finite pressure anisotropies as well as strong
toroidal and poloidal differential rotation, characteristic of Tokamak
plasmas. Despite the fact that physical phenomena occurring in fusion
plasmas depend fundamentally on the microscopic particle phase-space
dynamics, their consistent kinetic treatment remains still essentially
unchalleged to date. The goal of this paper is to address the problem within
the framework of Vlasov-Maxwell description. The gyrokinetic treatment of
charged particles dynamics is adopted for the construction of asymptotic
solutions for the quasi-stationary species kinetic distribution functions.
These are expressed in terms of the particle exact and adiabatic invariants.
The theory relies on a perturbative approach, which permits to construct
asymptotic analytical solutions of the Vlasov-Maxwell system. In this way,
both diamagnetic and energy corrections are included consistently into the
theory. In particular, by imposing suitable kinetic constraints, the
existence of generalized bi-Maxwellian asymptotic kinetic equilibria is
pointed out. The theory applies for toroidal rotation velocity of the order
of the ion thermal speed. These solutions satisfy identically also the
constraints imposed by the Maxwell equations, i.e. quasi-neutrality and
Ampere's law. As a result, it is shown that, in the presence of non-uniform
fluid and EM fields, these kinetic equilibria can sustain simultaneously
toroidal differential rotation, quasi-stationary finite poloidal flows and
temperature anisotropy.
\end{abstract}

\maketitle










\section{Introduction}

Plasma dynamics is most frequently treated in the framework of stand-alone
MHD approaches, i.e., formulated independent of an underlying kinetic
theory. However, these treatments can provide at most a partial description
of plasma phenomenology. The reason is related to two basic inconsistencies
of customary fluid approaches. First, the set of fluid equations may not be
closed, requiring in principle the prescription of arbitrary higher-order
fluid fields. Second, in these approaches typically no account is given of
microscopic phase-space particle dynamics as well as phase-space plasma
collective phenomena. It is well known that only in the context of kinetic
theory these difficulties can be consistently met. Such a treatment in fact
permits to obtain well-defined constitutive equations for the relevant fluid
fields describing the plasma state, overcoming at the same time the closure
problem. Kinetic theory is appropriate, for example, in the case of
collisionless or weakly-collisional plasmas where phase-space particle
dynamics is expected to play a dominant role.

Unfortunately, for a wide range of physical effects arising in
magnetically-confined plasmas and relevant for controlled fusion research, a
fully consistent approach of this type is still missing. Surprisingly, these
include even the description of equilibrium or slowly-time varying phenomena
occurring in realistic laboratory Tokamak plasmas. The issue concerns
specifically the description of finite pressure anisotropies, strong
toroidal differential rotation as well as concurrent poloidal flows observed
in Tokamak devices. The deficiency may represent a serious obstacle for
meaningful developments in plasma physics (both theoretical and
computational) and controlled fusion research. In particular, it is
well-known that both toroidal and poloidal plasma equilibrium rotation flows
may exist in Tokamak plasmas \cite{Eriksson1997,Zhou2010}. The observation
of intrinsic rotation, occurring without any external momentum source \cite%
{Boedo2011}, remains essentially unexplained to date, being mostly ascribed
to turbulence or boundary-layer phenomena occurring in the outer regions of
the plasma \cite{Catto2009,Pamela2010}. Such an effect, potentially
combining both toroidal and poloidal flow velocities with temperature
anisotropy, may be of critical importance both for stability and suppression
of turbulence \cite{Hameiri1983,Hassam96,Max1998}.

The goal of the present investigation is the construction of slowly-time
varying particular solutions of the Vlasov-Maxwell system for collisionless
axisymmetric plasmas immersed in strong magnetic and electric fields. In
principle, two approaches are possible for the investigation of the problem.
One is based on the Chapman-Enskog solution of the drift-kinetic Vlasov
equation, namely achieved by seeking a perturbative solution of the form $%
f_{s}=f_{Ms}+\varepsilon f_{1s}+...,$ where $0<\varepsilon \ll 1$ is an
appropriate dimensionless parameter to be defined below (see Section 2) and $%
f_{Ms}$ a suitable equilibrium kinetic distribution function (KDF). In
customary formulations this is typically identified with a drifted
Maxwellian KDF. An example is provided by Hinton \textit{et al.} \cite%
{Brizard1994} where an approximate equilibrium KDF carrying both toroidal
and poloidal flows was introduced to describe ion poloidal flows in Tokamaks
near the plasma edge. An alternative approach is represented by the
construction of exact or asymptotic solutions of the Vlasov equations of the
form $f_{s}=f_{\ast s},$ with $f_{\ast s}$ to be considered only function of
particle exact and adiabatic invariants, via the introduction of suitable
\textit{kinetic constraints}. This technique is exemplified by Ref.\cite%
{Catto1987}, where $f_{\ast s}$ was assumed to be a function of only two
invariants, namely the particle energy $E_{s}\equiv Z_{s}e\Phi _{\ast s}$
and toroidal canonical momentum $p_{\varphi s}\equiv \frac{Z_{s}e}{c}\psi
_{\ast s}$ [see their definitions given below], and identified with a
generalized Maxwellian distribution of the form%
\begin{equation}
f_{\ast s}=\frac{n_{\ast s}}{\pi ^{3/2}\left( 2T_{\ast s}/M_{s}\right) ^{3/2}%
}\exp \left\{ -\frac{H_{\ast s}}{T_{\ast s}}\right\} .  \label{kdf87}
\end{equation}%
Here $H_{\ast s}$ is the invariant $H_{\ast s}\equiv E_{s}-\frac{Z_{s}e}{c}%
\int\limits_{0}^{\psi _{\ast s}}d\psi \Omega _{0}(\psi ),$ while $\Lambda
_{\ast s}\equiv \left\{ n_{\ast s,}T_{\ast s}\right\} $\ denotes suitable
\textquotedblleft structure functions\textquotedblright , i.e., properly
defined functions of the particle invariants. In Refs.\cite%
{Catto1987,Max1992,Max1996} these were prescribed imposing the kinetic
constraint $\Lambda _{\ast s}=\Lambda _{\ast s}(\psi _{\ast s})$. By
performing a perturbative expansion in the canonical momentum (see also the
related discussion in Section 6), it was shown that $f_{\ast s}$ recovers
the Chapman-Enskog form, with the leading-order Maxwellian KDF carrying
isotropic temperature $T_{s}(\psi )$, species-independent toroidal angular
rotation velocity $\Omega _{0}(\psi )$ (see definition given by Eq.(\ref%
{omega0})) and finite toroidal differential rotation, i.e., $\frac{\partial
}{\partial \psi }\Omega _{0}(\psi )\neq 0$. A basic aspect of Tokamak
plasmas is the property of allowing toroidal rotation velocities $R\Omega
_{0}$ comparable to the ion thermal velocity $v_{thi}=\left\{
2T_{i}/M_{i}\right\} ^{1/2}$. As shown in Ref.\cite{Catto1987} this implies
the fundamental consequence that, for kinetic equilibria characterized by
purely toroidal differential rotation as described by the KDF (\ref{kdf87}),
necessarily the self-generated electrostatic (ES) potential $\Phi $ in the
plasma must satisfy the ordering $\left( M_{i}v_{thi}^{2}\right) /\left(
Z_{i}e\Phi \right) \sim O\left( \varepsilon \right) $. If the ion and
electron temperatures are comparable, in the sense that $T_{i}/T_{e}\sim
O\left( \varepsilon ^{0}\right) $, it follows that an analogous ordering
must hold also for the electron species. \textit{Therefore, the same
asymptotic condition must be adopted for all thermal particles of the
plasma, namely for which }$\left\vert \mathbf{v}\right\vert \sim v_{ths}$%
\textit{, independent of species.}

In the following, utilizing such a type of ordering, the second route is
adopted. Hence, the theory developed here applies to a two-species
ion-electron plasma characterized by toroidal rotation velocity of the order
of the ion thermal speed. It relies on the perturbative kinetic theory
developed in Refs.\cite{Cr2010,Cr2011} (hereafter referred to as Papers I
and II). The aim is to provide a systematic generalization of the theory
presented in Ref.\cite{Catto1987}, allowing $f_{\ast s}$ to depend on the
complete set of independent adiabatic invariants, and therefore to vary
slowly in time (\textquotedblleft equilibrium\textquotedblright\ KDF). In
particular, here we intend to show that, besides the properties indicated
above, also temperature anisotropy, finite poloidal flow velocities and
first-order perturbative corrections, including finite Larmor-radius (FLR)
corrections, can be consistently dealt with at the equilibrium level. A
remarkable feature of the approach is that, by construction, all the moment
equations stemming from the Vlasov equation are identically satisfied,
together with their related solubility conditions (i.e., following from the
condition of periodicity of the KDF and its moments in the poloidal angle).
An interesting development consists in the inclusion of both diamagnetic
(i.e., FLR) and energy corrections arising from the Taylor-expansions of the
relevant structure functions. In such a case the structure functions are
identified with smooth functions of both the particle energy and toroidal
canonical momentum, of the general form%
\begin{equation}
\Lambda _{\ast s}=\Lambda _{s}(\psi _{\ast s},\Phi _{\ast s}),
\label{kincon1}
\end{equation}%
with the functions $\Lambda _{s}(\psi ,\Phi )$ being identified with
suitable fluid fields, $s$ denoting the species index. This permits the
construction of a systematic perturbative expansion also for the KDF itself,
allowing to retain perturbative corrections (of arbitrary order) expressed
as polynomial functions in terms of the particle velocity. In particular,
under suitable assumptions, the leading-order KDF is shown to be determined
by a bi-Maxwellian distribution carrying anisotropic temperature and
non-uniform, both toroidal and poloidal, flow velocities. Thanks to the
kinetic constraints, constitutive equations are determined for the related
equilibrium fluid fields. First-order corrections with respect to $%
\varepsilon $ are shown to be linear functions of suitably-generalized
thermodynamic forces. These include now, besides the customary ones \cite%
{Catto1987}, additional thermodynamic forces associated to energy
derivatives of the relevant structure functions.

The constraints imposed by the Maxwell equations are then investigated.
First, the Poisson equation is analyzed within the quasi-neutrality
approximation. As a development with respect to Ref.\cite{Catto1987}, it is
proved that the perturbative scheme determines uniquely, correct through $%
O\left( \varepsilon ^{0}\right) $, the equilibrium ES potential, including
the $1/O\left( \varepsilon \right) $ contribution. Second, the solubility
conditions of Ampere's law are shown to prescribe constraints on the species
poloidal and toroidal flow velocities and the corresponding current
densities. The theory applies for magnetic configurations with nested and
closed toroidal magnetic surfaces characterized by finite aspect ratio.

The paper is organized as follows. First, in Sections 2 and 3 the
Vlasov-Maxwell and magnetized-plasma asymptotic orderings are posed,
together with the basic assumptions concerning the plasma and its
electromagnetic (EM) field. In Section 4 particle first integrals and
adiabatic invariants are recalled, including guiding-center adiabatic
invariants predicted by gyrokinetic theory. In Section 5 particular
solutions of the collisionless Vlasov equation are investigated. Their
representation in terms of suitable structure functions is discussed. Then,
based on the Taylor-expansion of the relevant structure functions, in
Section 6 a perturbative kinetic theory is obtained for the KDF. As an
application, the leading-order and the first-order diamagnetic and energy
contributions to the KDF are displayed. In Section 7 the connection between
the kinetic and fluid treatments is addressed. In Section 8 the
leading-order number density and flow velocity carried by the stationary KDF
are reported. The implications of Maxwell equations are discussed in
Sections 9 and 10. In particular, in Section 9 the issue concerning the
quasi-neutrality condition is addressed. It is shown that quasi-neutrality
determines uniquely, up to an arbitrary constant, the ES potential (THM.1)
and is consistent with the plasma asymptotic orderings introduced (Corollary
to THM.1). Then, constraints placed by the Ampere equation are investigated
in Section 10 (THM.2). Relevant comparisons with previous literature, based
either on kinetic or fluid approaches, are presented in Section 11. Finally,
concluding remarks are given in Section 12.

\section{Vlasov-Maxwell asymptotic orderings}

In the following, for particles belonging to the $s$-species, we introduce
the characteristic time and length scales $\Delta t_{s}\equiv \frac{2\pi r}{%
v_{\perp ths}}$ and $\Delta L_{s}=\Delta L\equiv 2\pi r,$ with $2\pi r$ and $%
v_{\perp ths}=\left\{ T_{\perp s}/M_{s}\right\} ^{1/2}$ denoting
respectively the \textit{connection length }and the thermal velocity
associated to the species perpendicular temperature $T_{\perp s}$ (defined
with respect to the local magnetic field direction). We shall consider
phenomena occurring in time intervals $\Delta t_{s}$ which belong to the
ranges $\tau _{ps}\ll \Delta t_{s}\ll \tau _{Cs},$ where $\tau _{ps}\equiv
\left( \frac{M_{s}}{4\pi n_{s}\left( Z_{s}e\right) ^{2}}\right) ^{1/2}$, and
for isotropic species temperatures $\tau _{Cs}\equiv \frac{3\sqrt{M_{s}}%
T_{s}^{3/2}}{4\sqrt{2\pi }n_{s}\ln \Lambda \left( Z_{s}e\right) ^{4}}$
denote respectively the Langmuir time and the Spitzer ion self-collision
time. A similar ordering follows for the corresponding scale-length $\Delta
L_{s}$ letting $\Delta L_{s}=\Delta t_{s}v_{ths}$, with $v_{ths}$ being the
species isotropic-temperature thermal velocity. For definiteness, we shall
consider here a plasma consisting of $n$ species of charged particles, with $%
n\geq 2$. Such a plasma can be regarded, respectively, as:

(\#1) \emph{Collisionless:} in validity of the inequality between $\Delta
t_{s}$ and $\tau _{Cs}$, contributions proportional to the ratio $%
\varepsilon _{Cs}\equiv \frac{\Delta t_{s}}{\tau _{Cs}}\ll 1$, here referred
to as the \textit{collision-time parameter}, can be ignored. Thus, Coulomb
binary interactions are negligible, so that all particle species in the
plasma can be regarded as collisionless.

(\#2) \emph{Continuous:} thanks to the left-side inequality between $\Delta
t_{s}$ and $\tau _{ps}$, plasma particles interact with each other only\ via
a continuum mean EM field. In particular, the inequality $\varepsilon
_{Lg,s}\equiv \frac{\tau _{ps}}{\Delta t_{s}}\ll 1$ is assumed to hold, with
$\varepsilon _{Lg,s}$ denoting the \textit{Langmuir-time parameter}.

(\#3) \emph{Quasi-neutral:} due again to the same inequality,\ the plasma is
quasi-neutral on the spatial scale $\Delta L_{s}$ corresponding to $\Delta
t_{s}$.

Systems fulfilling requirements \#1-\#2 - the so-called \emph{Vlasov-Maxwell
plasmas }-\emph{\ }rely on kinetic theory, since fluid MHD approaches are
inapplicable in such a case (see related discussion in Papers I and II).
Such plasmas are described in the framework of the Vlasov-Maxwell kinetic
theory. In this case the plasma is treated as an ensemble of particle $s-$%
species (subsets of like particles) each one described by a KDF $f_{s}(%
\mathbf{z},t)$ defined in the phase-space $\Gamma =\Gamma _{r}\times \Gamma
_{u}$ (with $\Gamma _{r}\subset
\mathbb{R}
^{3}$ and $\Gamma _{u}\equiv
\mathbb{R}
^{3}$ denoting respectively the configuration$\ $and velocity spaces) and
satisfying the Vlasov kinetic equation. Velocity moments of $f_{s}(\mathbf{z}%
,t)$ are then defined as integrals of the form $\int_{\Gamma _{u}}d^{3}vQ(%
\mathbf{z},t)f_{s}(\mathbf{z},t)$, with $Q(\mathbf{z},t)$ being a suitable
phase-space weight function. In particular, for $Q(\mathbf{z},t)=\left\{ 1,%
\mathbf{v}\right\} $ the velocity moments determine the source of the EM
self-field $\left\{ \mathbf{E}^{self},\mathbf{B}^{self}\right\} $,
identified with the plasma charge and current densities $\left\{ \rho (%
\mathbf{r},t),\mathbf{J}(\mathbf{r},t)\right\} $.

In addition, we require the plasma to be\ \emph{axisymmetric}, so that, when
referred to a set of cylindrical coordinates $(R,\varphi ,z),$ all relevant
dynamical variables characterizing the plasma (e.g., the fluid fields and
the EM field) are independent of the azimuthal angle $\varphi $. Here, by
assumption, the configuration space is identified with the bounded internal
domain of an axisymmetric torus, which can be parametrized in terms of the
scale-lengths $\left( a,R_{0}\right) $, with $a$ denoting $a\equiv \sup
\left\{ r,r\in \Gamma _{r}\right\} $ and $R_{0}$ the radius of the plasma
magnetic axis.

\section{Basic assumptions}

In this section the basic hypothesis of the model, which include the EM
field and the magnetized-plasma orderings, are pointed out.

\subsection{The EM field}

Here we restrict our analysis to EM fields which are slowly-time varying in
the sense $\left[ \mathbf{E}(\mathbf{x},\varepsilon ^{k}t),\mathbf{B}(%
\mathbf{x},\varepsilon ^{k}t)\right] $, with $k\geq 1$ being a suitable
integer (\emph{quasi-stationarity condition}). This type of time dependence
is thought to arise either due to external sources or boundary conditions.
In particular, the magnetic field $\mathbf{B}$ is assumed to be of the form%
\begin{equation}
\mathbf{B}\equiv \nabla \times \mathbf{A}=\mathbf{B}^{self}(\mathbf{x}%
,\varepsilon ^{k}t)+\mathbf{B}^{ext}(\mathbf{x},\varepsilon ^{k}t),
\label{b1}
\end{equation}%
where $\mathbf{B}^{self}$ and $\mathbf{B}^{ext}$ denote the self-generated
magnetic field produced by the plasma and a finite external magnetic field
produced by external coils.\ In particular the magnetic field $\mathbf{B}$
admits by assumption a family of nested and closed\textit{\ }axisymmetric
toroidal magnetic surfaces $\left\{ \psi (\mathbf{\ x})\right\} \equiv
\left\{ \psi (\mathbf{x})=const.\right\} $, where $\psi $ denotes the
poloidal magnetic flux of $\mathbf{B}$ and, because of axisymmetry, $\mathbf{%
x}$ can be identified with the coordinates $\mathbf{x}=\left( R,z\right) $.
In such a setting a set of magnetic coordinates ($\psi ,\varphi ,\vartheta $%
) can be defined, where $\vartheta $ is a curvilinear angle-like coordinate
on the magnetic surfaces $\psi (\mathbf{x})=const.$ It is assumed that the
vectors ($\nabla \psi ,\nabla \varphi ,\nabla \vartheta $) define a
right-handed system. Each relevant physical quantity $G(\mathbf{x},t)$ can
then be conveniently expressed either in terms of the cylindrical
coordinates or as a function of the magnetic coordinates, i.e. $G(\mathbf{x}%
,t)=\overline{G}\left( \psi ,\vartheta ,t\right) $. The total magnetic field
is then decomposed as%
\begin{equation}
\left. \mathbf{B}=I(\mathbf{x},\varepsilon ^{k}t)\nabla \varphi +\nabla \psi
(\mathbf{x},\varepsilon ^{k}t)\times \nabla \varphi ,\right.   \label{bself}
\end{equation}%
where $\mathbf{B}_{T}\equiv I(\mathbf{x},\varepsilon ^{k}t)\nabla \varphi $
and $\mathbf{B}_{P}\equiv \nabla \psi (\mathbf{x},\varepsilon ^{k}t)\times
\nabla \varphi $ are the toroidal and poloidal components of the field. In
particular, the following ordering is assumed to hold: $\frac{\left\vert
\mathbf{B}_{P}\right\vert }{\left\vert \mathbf{B}_{T}\right\vert }\sim
O\left( \varepsilon ^{0}\right) .$ Finally, the corresponding electric field
expressed in terms of the EM potentials $\left\{ \Phi (\mathbf{x}%
,\varepsilon ^{k}t),\mathbf{A}(\mathbf{x},\varepsilon ^{k}t)\right\} $ is
considered primarily electrostatic, namely%
\begin{equation}
\mathbf{E}(\mathbf{x},\varepsilon ^{k}t)\equiv -\nabla \Phi -\varepsilon ^{k}%
\frac{1}{c}\frac{\partial \mathbf{A}}{\partial \tau }\cong -\nabla \Phi ,
\label{electric}
\end{equation}%
with $\tau $ denoting the slow-time variable $\tau \equiv \varepsilon ^{k}t$%
, and quasi-orthogonal to the magnetic field, in the sense that $\frac{%
\mathbf{E}\cdot \mathbf{B}}{\left\vert \mathbf{E}\right\vert \left\vert
\mathbf{B}\right\vert }\sim O\left( \varepsilon \right) $, while $\frac{%
c\left\vert \mathbf{E}\right\vert }{\left\vert \mathbf{B}\right\vert }\frac{1%
}{v_{ths}}\sim O\left( \varepsilon ^{0}\right) .$ Together with the
quasi-stationarity condition, this implies that, to leading order in $%
\varepsilon ,$ $\Phi =\Phi (\psi ,\varepsilon ^{k}t)$. In particular,
assuming that both $\Phi $ and $\mathbf{A}$ are analytic with respect to $%
\varepsilon $, it can be shown that, consistent with gyrokinetic (GK) theory
and the asymptotic orderings indicated below (see next Section), they must
be considered of the general form%
\begin{eqnarray}
\Phi  &=&\frac{1}{\varepsilon }\Phi _{-1}\left( \psi ,\varepsilon
^{k}t\right) +\varepsilon ^{0}\Phi _{0}\left( \psi ,\vartheta ,\varepsilon
^{k}t\right) +..,  \label{Laurent} \\
\mathbf{A} &=&\frac{1}{\varepsilon }\mathbf{A}_{-1}\left( \mathbf{r}%
,\varepsilon ^{k}t\right) +\varepsilon ^{0}\mathbf{A}_{0}\left( \mathbf{r}%
,\varepsilon ^{k}t\right) +..,  \label{Laurent2}
\end{eqnarray}%
where $\Phi $ is expressed in terms of the magnetic coordinates and $\mathbf{%
A}_{-1}$ is $\mathbf{A}_{-1}\equiv \psi \nabla \varphi +g\left( \psi
,\vartheta ,\varepsilon ^{k}t\right) \nabla \vartheta $, with $g$ being a
suitable function.

\subsection{The magnetized-plasma orderings}

Next, let us introduce the \textit{magnetized plasma ordering} appropriate
for the treatment of single-particle dynamics in magnetized plasmas, i.e.
for which in particular $B^{2}\gg E^{2}$. For $s=i,e$, this requires the
definition of the following additional dimensionless parameters:

1) \textit{Larmor-radius parameter }$\varepsilon _{M,s}\equiv \frac{r_{Ls}}{%
\Delta L_{s}}$ and \textit{Larmor-time parameter }$\varepsilon _{Lr,s}\equiv
\frac{\tau _{Ls}}{\Delta t_{s}}:$ here $\tau _{Ls}$ and $r_{Ls}$ are
respectively the Larmor time and the Larmor radius of the species $s$, with $%
s=1,n$, defined as $r_{Ls}\equiv v_{\perp ths}/\Omega _{cs}$, with $\Omega
_{cs}=Z_{s}eB/M_{s}c\equiv 1/\tau _{Ls}$ being the species Larmor frequency.
Imposing the requirement that $\tau _{Ls}\ll \Delta t_{s}$ and $r_{Ls}\ll
\Delta L_{s},$ it follows that $\varepsilon _{M,s}$ and $\varepsilon _{Lr,s}$
are infinitesimals of the same order, i.e., $0\leq \varepsilon _{Lr,s}\sim
\varepsilon _{M,s}\ll 1.$ Requiring again that $T_{i}\sim T_{e}$, and
furthermore $Z_{i}\sim O\left( 1\right) $, it follows that $\varepsilon
_{M,i}\sim \left( \frac{M_{i}}{M_{e}}\right) ^{1/2}\varepsilon _{M,e}$.

2) \textit{Canonical-momentum parameter: }$\varepsilon _{s}\equiv \left\vert
\frac{L_{\varphi s}}{p_{\varphi s}-L_{\varphi s}}\right\vert =\left\vert
\frac{M_{s}Rv_{\varphi }}{\frac{Z_{s}e}{c}\psi }\right\vert $, where $%
v_{\varphi }\equiv \mathbf{v}\cdot \mathbf{e}_{\varphi }$ and $L_{\varphi s}$
denotes the species particle angular momentum.

3) \textit{Total-energy parameter: }$\sigma _{s}\equiv \left\vert \frac{%
\frac{M_{s}}{2}v^{2}}{{Z_{s}e}\Phi }\right\vert $, where $\frac{M_{s}}{2}%
v^{2}\sim T_{s}$ and ${Z_{s}e}\Phi $ are respectively the particle kinetic
and ES energy.

In principle, the parameters $\varepsilon _{s}$ and $\sigma _{s}$ are
independent (in particular, as pointed out in Paper I, they might differ
from $\varepsilon _{M,s}$). More precisely, here we shall consider the
subset of phase-space for which the following ordering holds:%
\begin{equation}
0\leq \sigma _{s}\sim \varepsilon _{s}\sim \varepsilon _{Lr,s}\sim
\varepsilon _{M,s}\ll 1,  \label{ordnow}
\end{equation}%
which applies in the subset of thermal particles. Notice that the assumption
on $\varepsilon _{s}$ is consistent with the requirement of finite inverse
aspect-ratio (see below), while, as recalled above, the ordering on $\sigma
_{s}$ is required for the treatment of Tokamak equilibria in the presence of
strong toroidal differential rotation \cite%
{Catto1987,Max1992,Max1996,Max1997}. The same orderings are of course
invoked also for the validity of the GK theory (see Ref.\cite{Briz1995} and
also Eq.(\ref{trGK}) in the next section and the related discussion). The
assumption on the $\sigma _{s}$-ordering can be shown to be consistent with
the quasi-neutrality condition (see Corollary to THM.1 in Section 9). The
previous requirements imply, for all species, the asymptotic perturbative
expansions in the variables $\psi _{\ast s}$ and $\Phi _{\ast s}$:%
\begin{eqnarray}
\psi _{\ast s} &=&\psi \left[ 1+O\left( \varepsilon _{M,s}\right) \right] ,
\label{psi} \\
\Phi _{\ast s} &=&\Phi \left[ 1+O\left( \varepsilon _{M,s}\right) \right] .
\label{es}
\end{eqnarray}

Finally, to warrant the validity of the Vlasov equation on the Larmor-radius
scale, we shall impose also that $0\ll \varepsilon _{mfp,s}\sim \varepsilon
_{Cs}\leq \varepsilon _{M,s},$ with $\varepsilon _{mfp,s}\equiv \frac{\Delta
L}{\lambda _{Cs}}$ and $\varepsilon _{Cs}\equiv \frac{\Delta t_{s}}{\tau
_{Cs}}$\textit{\ }denoting respectively the \textit{mean-free-path parameter
}and\textit{\ the collision-time parameter. }Then, consistent with
quasi-neutrality, we demand also that $\varepsilon _{Lg,s}\sim \varepsilon
_{D}\leq \varepsilon _{M,s}\leq \varepsilon \ll 1,$ with $\varepsilon =\sup
\left\{ \varepsilon _{M,s},\text{ }s=e,i\right\} .$ Finally, the \textit{%
inverse aspect-ratio parameter} $\delta \equiv \frac{a}{R_{0}}$ will be
considered finite, i.e. such that $\delta \sim O\left( \varepsilon
^{0}\right) .$ We remark that the parameters $\left\{ \sigma
_{s},\varepsilon _{s},\varepsilon _{Lr,s},\varepsilon _{M,s}\right\} $ deal
with the single-particle dynamics, $\left\{ \varepsilon _{Lg,s},\varepsilon
_{D,s},\varepsilon _{mfp,s},\varepsilon _{Cs}\right\} $ concern collective
properties of the plasma, while $\delta $ is a purely geometrical quantity.

\section{The particle adiabatic invariants}

For single-particle dynamics, the exact first integrals of motion and the
relevant adiabatic invariants are well-known. In particular, the adiabatic
invariants can be defined either in the context of Hamiltonian dynamics or
GK theory \cite{Bern1985,Little1979,Little1981}. In both cases, for a
magnetized plasma, they can be referred to the Larmor frequency. Hence, by
definition, a phase-function $P_{s}$ depending on the $s$-species particle
state is denoted as adiabatic invariant of order $n$ with respect to $%
\varepsilon _{M,s}$ if it is conserved asymptotically, namely in the sense $%
\frac{1}{\Omega _{cs}^{\prime }}\frac{d}{dt}\ln P_{s}=0+O(\varepsilon
_{M,s}^{n+1})$, where $n\geq 0$ is a suitable integer and $\Omega
_{cs}^{\prime }$ is the Larmor frequency evaluated at the guiding-center
position $\mathbf{x}^{\prime }$. Note that, in the following, we shall use a
prime \textquotedblleft\ $^{\prime }$ \textquotedblright\ to denote a
dynamical variable defined at the \emph{guiding-center position} $\mathbf{r}%
^{\prime }$ (or $\mathbf{x}^{\prime }$ in axisymmetry). Under the
assumptions of axisymmetry, the only first integral of motion is the
canonical momentum $p_{\varphi s}$ conjugate to the azimuthal angle $\varphi
$:%
\begin{equation}
p_{\varphi s}=M_{s}R\mathbf{v\cdot e}_{\varphi }+\frac{Z_{s}e}{c}\psi \equiv
\frac{Z_{s}e}{c}\psi _{\ast s}.  \label{p_fi}
\end{equation}%
Furthermore, the total particle energy%
\begin{equation}
E_{s}=\frac{M_{s}}{2}v^{2}\mathbf{+}{Z_{s}e}\Phi (\mathbf{x},\varepsilon
^{n}t)\equiv Z_{s}e\Phi _{\ast s},  \label{total_energy}
\end{equation}%
with $n\geq 1,$ is assumed to be an adiabatic invariant of order $n$.

Let us now analyze the adiabatic invariants predicted by GK theory. As
usual, the GK treatment involves the construction - in terms of an
asymptotic perturbative expansion determined by means of a power series in $%
\varepsilon _{M,s}$ - of a diffeomorphism of the form $\mathbf{z}\equiv (%
\mathbf{\mathbf{r}},\mathbf{v})\mathbf{\rightarrow z}^{\prime }\equiv (%
\mathbf{r}^{\prime },\mathbf{v}^{\prime })$, referred to as the \emph{GK
transformation}. The GK transformation is performed on all phase-space
variables $\mathbf{z}\equiv (\mathbf{\mathbf{r}},\mathbf{v})$, \textit{except%
} for the azimuthal angle $\varphi $ which is left unchanged and is
therefore to be considered as one of the GK variables. Here, by definition,
the transformed variables $\mathbf{z}^{\prime }$\textbf{\ }(\emph{GK state})
are constructed so that their time derivatives to the relevant order in $%
\varepsilon _{M,s}$ have at least one ignorable coordinate, to be identified
with a suitably-defined gyrophase $\phi ^{\prime }$. Starting point is then
the representation of the particle Lagrangian in terms of the hybrid
variables $\mathbf{z}$. This is expressed as $\mathcal{L}_{s}(\mathbf{z},%
\frac{d}{dt}\mathbf{z},\varepsilon ^{k}t)\equiv \mathbf{\dot{r}}\cdot
\mathbf{P}_{s}-\mathcal{H}_{s}(\mathbf{z},\varepsilon ^{k}t),$ where $%
\mathbf{P}_{s}\equiv \left[ M_{s}\mathbf{v}+\frac{Z_{s}e}{c}\mathbf{A}(%
\mathbf{x},\varepsilon ^{k}t)\right] $ and $\mathcal{H}_{s}(\mathbf{z}%
,\varepsilon ^{k}t)=\frac{M_{s}}{2}v^{2}\mathbf{+}{Z_{s}e}\Phi (\mathbf{x}%
,\varepsilon ^{k}t)$ denotes the corresponding Hamiltonian function in
hybrid variables. The development of GK theory is well known. It involves a
phase-space transformation to a local reference frame in which the particle
guiding-center is instantaneously at rest with respect to the $\psi $%
-surface to which it belongs. In this case, the leading-order GK
transformation can be proved to be necessarily of the form%
\begin{equation}
\left\{
\begin{array}{c}
\mathbf{r}=\mathbf{r}^{\prime }-\frac{\mathbf{w}^{\prime }\times \mathbf{b}%
^{\prime }}{\Omega _{cs}^{\prime }}, \\
\mathbf{v}=u^{\prime }\mathbf{b}^{\prime }+\mathbf{w}^{\prime }+\mathbf{U}%
^{\prime },%
\end{array}%
\right.   \label{trGK}
\end{equation}%
Here, in particular, $\mathbf{U}^{\prime }\equiv \mathbf{U}(\mathbf{x}%
^{\prime },\varepsilon ^{k}t)$, with $\mathbf{U}(\mathbf{x},\varepsilon
^{k}t)$ being the fluid-field identified with the $\mathbf{E}\times \mathbf{B%
}-$drift velocity:%
\begin{equation}
\mathbf{U}(\mathbf{x},\varepsilon ^{k}t)\equiv -\frac{c}{{B}}\nabla \Phi
\times \mathbf{b}.  \label{u}
\end{equation}%
This coincides with the so-called frozen-in velocity, namely the fluid
velocity with respect to which each line of force is carried into itself.
The rest of the notation is standard. Thus, $u^{\prime }$ and $\mathbf{w}%
^{\prime }$ denote respectively the parallel and perpendicular
(guiding-center) velocities, with $\mathbf{w}^{\prime }=w^{\prime }\cos \phi
^{\prime }\mathbf{e}_{1}^{\prime }+w^{\prime }\sin \phi ^{\prime }\mathbf{e}%
_{2}^{\prime }$ and $\phi ^{\prime }$ denoting the gyrophase angle, $\Omega
_{cs}^{\prime }=\frac{Z_{s}eB^{\prime }}{M_{s}c}$ and $\mathbf{b}^{\prime }=%
\mathbf{b}(\mathbf{x}^{\prime },\varepsilon ^{k}t),$ with $\mathbf{b}(%
\mathbf{x},\varepsilon ^{k}t)\mathbf{\equiv B}($\textbf{$\mathbf{x}$}$%
,\varepsilon ^{k}t)\mathbf{/}B(\mathbf{x},\varepsilon ^{k}t)$. Notice that,
here, by construction, $\left\vert \frac{\mathbf{w}^{\prime }\times \mathbf{b%
}^{\prime }}{\Omega _{cs}^{\prime }}\right\vert $ must be considered of $%
O\left( \varepsilon _{M,s}\right) $ with respect to $\left\vert \mathbf{r}%
^{\prime }\right\vert $, while for thermal particles $\left\vert u^{\prime
}\right\vert $ and $\left\vert \mathbf{w}^{\prime }\right\vert $ are all of
the same order of $v_{ths}$. In particular, due to the previous orderings,
for the validity of GK theory the EM potentials $\left( \Phi ,\mathbf{A}%
\right) $ entering the Lagrangian must be considered of the form indicated
above (see Eqs.(\ref{Laurent}) and (\ref{Laurent2})), namely both of $%
1/O\left( \varepsilon \right) $ with respect to the remaining terms. As a
consequence, the ordering (\ref{ordnow}) for $\sigma _{s}$ necessarily
applies, under the assumption $T_{i}/T_{e}\sim O\left( \varepsilon
^{0}\right) $ considered here. On the other hand, as in Ref.\cite{Catto1987}%
, $\left\vert \mathbf{U}^{\prime }\right\vert $ is to be taken of the order
of the ion thermal velocity $v_{thi}$, while $\left\vert \mathbf{U}^{\prime
}\right\vert \sim \Omega _{0}R$, with $\Omega _{0}$ being the\ toroidal
angular rotation frequency, defined below by Eq.(\ref{omega0}). \textit{It
is important to stress here that these two conditions imply that }$\Phi $%
\textit{\ must satisfy the asymptotic ordering given above by Eq.(\ref%
{Laurent}). Therefore, the previous orderings for }$\sigma _{s}$\textit{\
and }$\Phi $\textit{\ must be regarded as basic prerequisites for the
description of Tokamak plasmas characterized by toroidal rotation speeds
comparable to the ion thermal velocity.}

By construction, in GK description the gyrophase angle is ignorable, so that
the magnetic moment $m_{s}^{\prime }$ is an adiabatic invariant of
prescribed accuracy. In particular, the leading-order approximation is $%
m_{s}^{\prime }\cong \mu _{s}^{\prime }\equiv \frac{M_{s}w^{\prime 2}}{%
2B^{\prime }}$. Two further adiabatic invariants can immediately be obtained
from the previous considerations. In fact, since the azimuthal angle $%
\varphi $ is ignorable also in GK theory, the conjugate GK canonical
momentum $p_{\varphi s}^{\prime }$, referred to as the \textit{%
guiding-center canonical momentum}, is necessarily an adiabatic invariant.
Neglecting corrections of $O(\varepsilon _{M,s})$ this is given by%
\begin{equation}
p_{\varphi s}^{\prime }\equiv \frac{M_{s}}{B^{\prime }}\left( u^{\prime
}I^{\prime }+\frac{c\nabla ^{\prime }\psi ^{\prime }\cdot \nabla ^{\prime
}\Phi ^{^{\prime }}}{B^{\prime }}\right) +\frac{Z_{s}e}{c}\psi ^{\prime },
\label{p_fi-HAT}
\end{equation}%
which provides a third-order adiabatic invariant. We remark that both $%
m_{s}^{\prime }$ and $p_{\varphi s}^{\prime }$ can in principle be
identified with adiabatic invariants of $O(\varepsilon _{M,s}^{k+1})$, with $%
k\geq 1$ arbitrarily prescribed \cite{Kruskal}. In the following we shall
make use of the local invariants $\left( \psi _{\ast s},E_{s},m_{s}^{\prime
}\right) $ to represent the particle state, while adopting $p_{\varphi
s}^{\prime }$ to deal with the dependences in terms of $u^{\prime }$.

\bigskip

\section{Vlasov kinetic theory: equilibrium KDF}

Let us now proceed constructing asymptotic solutions of the Vlasov equation
holding for collisionless Tokamak plasmas in validity of the previous
assumptions. The treatment is based on Papers I and II, where equilibrium
generalized bi-Maxwellian solutions for the KDF were proved to hold for
accretion disk plasmas. In particular, the following features are required
for the equilibrium KDF:

1) For all of the species, different parallel and perpendicular temperatures
are allowed (temperature anisotropy).

2) Non-vanishing species dependent differential toroidal and poloidal
rotation velocities are included.

3) The KDF is required to be an adiabatic invariant asymptotically
\textquotedblleft close\textquotedblright\ to a local bi-Maxwellian. Hence,
in particular, in the case of a locally non-rotating plasma (i.e., for which
both toroidal and poloidal rotation velocities vanish identically on a given
$\psi $-surface) the KDF must be close to a locally non-rotating
bi-Maxwellian.

In analogy with Papers I and II, it is possible to show that Requirements 1)
- 3) can be fulfilled by a suitable modified bi-Maxwellian expressed solely
in terms of first integrals and adiabatic invariants, including also a
suitable set of structure functions $\left\{ \Lambda _{\ast s}\right\} $ of
the form (\ref{kincon1}) (see the precise definition below). Hence, the
desired KDF is identified with an adiabatic invariant of the form%
\begin{equation}
f_{\ast s}=f_{\ast s}\left( E_{s},\psi _{\ast s},p_{\varphi s}^{\prime
},m_{s}^{\prime },(\psi _{\ast s},\Phi _{\ast s}),\varepsilon ^{n}t\right) ,
\label{form}
\end{equation}%
with $n\geq 1$, and where the brackets $(\psi _{\ast s},\Phi _{\ast s})$
denote the dependence in terms of the structure functions $\left\{ \Lambda
_{\ast s}(\Phi _{\ast s},\psi _{\ast s})\right\} $. In particular, in
agreement with assumptions 1) - 3), $f_{\ast s}$ is identified with KDF of
the form:%
\begin{equation}
f_{\ast s}=\frac{\beta _{\ast s}}{\left( 2\pi /M_{s}\right) ^{3/2}\left(
T_{\parallel \ast s}\right) ^{1/2}}\exp \left\{ -\frac{E_{\ast s}}{%
T_{\parallel \ast s}}-m_{s}^{\prime }\alpha _{\ast s}\right\} ,  \label{sol2}
\end{equation}%
which we refer here to as the \emph{generalized bi-Maxwellian KDF with
parallel velocity perturbations}. The notation is as follows. First, $%
\left\{ \Lambda _{\ast s}\right\} \equiv \left\{ \beta _{\ast s},\alpha
_{\ast s},T_{\parallel \ast s},\Omega _{\ast s},\xi _{\ast s}\right\} $ are
structure functions subject to kinetic constraints of the type (\ref{kincon1}%
), assumed analytic functions of both $\psi _{\ast s}$ and $\Phi _{\ast s}.$
These are, by definition, suitably close to appropriate fluid fields $%
\Lambda _{s}=\Lambda _{s}(\psi ,\Phi )$. In particular, the functions $%
\Lambda _{s}$ are defined as $\left\{ \Lambda _{s}\right\} \equiv \left\{
\beta _{s}\equiv \frac{\eta _{s}}{T_{\perp s}},\alpha _{s}\equiv \frac{%
B^{\prime }}{\Delta _{T_{s}}},T_{\parallel s},\Omega _{s},\xi _{s}\right\} $%
, where $\eta _{s}$ denotes the \textit{pseudo-density}, $T_{\parallel s}$
and $T_{\perp s}$ the parallel and perpendicular temperatures, with $\frac{1%
}{\Delta _{T_{s}}}\equiv \frac{1}{T_{\perp s}}-\frac{1}{T_{\parallel s}}$,
while $\Omega _{s}$ and $\xi _{s}$ are the toroidal and parallel rotation
frequencies. Second, the phase-function $E_{\ast s}$ is defined as $E_{\ast
s}\equiv H_{\ast s}-p_{\varphi s}^{\prime }\xi _{\ast s},$ while $H_{\ast s}$
is identified with
\begin{equation}
H_{\ast s}\equiv E_{s}-\frac{Z_{s}e}{c}\psi _{\ast s}\Omega _{\ast s}.
\label{H*}
\end{equation}

We stress that the form of $f_{\ast s}$ [see Eq.(\ref{sol2})] is obtained
consistent with assumption 3), namely such that when the constraint $\Omega
_{\ast s}=\xi _{\ast s}=0$ \emph{locally holds}, $f_{\ast s}$ reduces to the
\emph{non-rotating generalized bi-Maxwellian KDF} $f_{\ast s}=\frac{\beta
_{\ast s}}{\left( 2\pi /M_{s}\right) ^{3/2}\left( T_{\parallel \ast
s}\right) ^{1/2}}\exp \left\{ -\frac{E_{s}}{T_{\parallel \ast s}}%
-m_{s}^{\prime }\alpha _{\ast s}\right\} .$ In particular, unlike Ref.\cite%
{Catto1987}, the definition given above for $H_{\ast s}$ follows by
requiring that $E_{\ast s}$, and hence also $H_{\ast s}$, is a \emph{local}
linear function of the frequencies $\Omega _{\ast s}$ and $\xi _{\ast s}$
and of the canonical momenta $p_{\varphi s}$ and $p_{\varphi s}^{\prime }.$

An equivalent representation for (\ref{sol2}) can be obtained invoking the
previous definitions. This yields:%
\begin{eqnarray}
&&\left. f_{\ast s}=\frac{\beta _{\ast s}\exp \left[ \frac{X_{\ast s}}{%
T_{\parallel \ast s}}\right] }{\left( 2\pi /M_{s}\right) ^{3/2}\left(
T_{\parallel \ast s}\right) ^{1/2}}\right.  \label{sol3} \\
&&\times \exp \left\{ -\frac{M_{s}\left( \mathbf{v}-\mathbf{W}_{\ast
s}-U_{\parallel \ast s}^{\prime }\mathbf{b}^{\prime }\right) ^{2}}{%
2T_{\parallel \ast s}}-m_{s}^{\prime }\alpha _{\ast s}\right\} ,  \notag
\end{eqnarray}%
where $\mathbf{W}_{\ast s}=\mathbf{e}_{\varphi }R\Omega _{\ast s},$ $%
U_{\parallel \ast s}^{\prime }=\frac{I^{\prime }}{B^{\prime }}\xi _{\ast s}$
and%
\begin{eqnarray}
X_{\ast s} &\equiv &M_{s}\frac{\left\vert \mathbf{W}_{\ast s}\right\vert ^{2}%
}{2}+\frac{Z_{s}e}{c}\psi \Omega _{\ast s}-Z_{s}e\Phi +\Upsilon _{\ast
s}^{\prime },  \label{FUNCTION X} \\
\Upsilon _{\ast s}^{\prime } &\equiv &\frac{M_{s}U_{\parallel \ast
s}^{\prime 2}}{2}\left( 1+\frac{2\Omega _{\ast s}}{\xi _{\ast s}}\right) +
\notag \\
&&+\left( \frac{M_{s}c\nabla ^{\prime }\psi ^{\prime }\cdot \nabla ^{\prime
}\Phi ^{\prime }}{B^{\prime 2}}+\frac{Z_{s}e}{c}\psi ^{\prime }\right) \xi
_{\ast s}.  \label{FINAL-1}
\end{eqnarray}%
Note that $U_{\parallel \ast s}^{\prime }$ is non-zero only if the toroidal
magnetic field is non-vanishing.

The following comments are in order:

1) $f_{\ast s}$ is by construction a solution of the \textit{asymptotic
Vlasov equation}%
\begin{equation}
\frac{1}{\Omega _{cs}^{\prime }}\frac{d}{dt}\ln f_{\ast s}=0+O\left(
\varepsilon ^{n+1}\right) .  \label{asvla}
\end{equation}

2) $f_{\ast s}$ is defined in the phase-space $\Gamma =\Gamma _{r}\times
\Gamma _{u}$, where $\Gamma _{r}$ and $\Gamma _{u}$ are both identified with
suitable subsets of the Euclidean space $%
\mathbb{R}
^{3}$. In particular, $f_{\ast s}$ is non-zero in the subset of phase-space
where the adiabatic invariants $p_{\varphi s}^{\prime }$, $\mathcal{H}%
_{s}^{\prime }$ and $m_{s}^{\prime }$ are defined. It follows that $f_{\ast
s}$ is suitable for describing both circulating and trapped particles.

3) The velocity moments of $f_{\ast s}$, to be identified with the
corresponding fluid fields, are unique once $f_{\ast s}$ is prescribed in
terms of the structure functions.

\section{Perturbation theory}

In this section we develop a perturbative kinetic theory for the KDF $%
f_{\ast s}$. This is obtained by performing on $f_{\ast s}$ a double-Taylor
expansion for the implicit functional dependences in the variables $\psi
_{\ast s}$ and $\Phi _{\ast s}$ \emph{carried only by the structure functions%
} $\left\{ \Lambda _{\ast s}\right\} $, while leaving unchanged all the
remaining phase-space dependences. As indicated above, such asymptotic
expansions can be expressed in terms of the dimensionless parameters $\sigma
_{s}$ and $\varepsilon _{s}$ in validity of the ordering (\ref{ordnow}).
Hence the double Taylor expansion yields:%
\begin{eqnarray}
\Lambda _{\ast s} &\cong &\Lambda _{s}+\left( \psi _{\ast s}-\psi \right)
\left[ \frac{\partial \Lambda _{\ast s}}{\partial \psi _{\ast s}}\right]
_{\substack{ \psi _{\ast s}=\psi  \\ \Phi _{\ast s}=\Phi }}+  \notag \\
&&+\left( \Phi _{\ast s}-\Phi \right) \left[ \frac{\partial \Lambda _{\ast s}%
}{\partial \Phi _{\ast s}}\right] _{\substack{ \psi _{\ast s}=\psi  \\ \Phi
_{\ast s}=\Phi }}+..,  \label{espan0}
\end{eqnarray}%
where both $\Lambda _{s}$ and the partial derivatives in (\ref{espan0}) are
by construction functions depending only on $\left( \psi ,\Phi \right) $.
This implies also their general dependence in terms of the magnetic
coordinates $\left( \psi ,\vartheta \right) $ (see Section 9). We notice
that the asymptotic order of the\ \textquotedblleft
gradients\textquotedblright\ of the structure functions $\frac{\partial
\Lambda _{\ast s}}{\partial \psi _{\ast s}}$ and $\frac{\partial \Lambda
_{\ast s}}{\partial \Phi _{\ast s}}$ depends whether in $\Lambda _{\ast s},$
$\psi _{\ast s}$ and/or $\Phi _{\ast s}$ are considered \textquotedblleft
fast\textquotedblright\ or \textquotedblleft slow\textquotedblright\
variables with respect to $\varepsilon ,$ in the sense that the same
gradients can be considered respectively $O(\varepsilon ^{0})$ or $%
O(\varepsilon )$. In principle, different possible orderings are allowed for
the perturbative expansion of $f_{\ast s}.$ Here we shall assume in
particular that the structure functions $\beta _{\ast s},\alpha _{\ast
s},T_{\parallel \ast s}$ have fast dependences, while $\Omega _{\ast s},\xi
_{\ast s}$ have only slow ones. As a consequence, the set of derivatives $%
\left\{ \frac{\partial \Omega _{\ast s}}{\partial \psi _{\ast s}},\frac{%
\partial \Omega _{\ast s}}{\partial \Phi _{\ast s}}\right\} $ and $\left\{
\frac{\partial \xi _{\ast s}}{\partial \psi _{\ast s}},\frac{\partial \xi
_{\ast s}}{\partial \Phi _{\ast s}}\right\} $ are both taken here as $%
O\left( \varepsilon \right) $. It follows that to first order in $%
\varepsilon $ the KDF $f_{\ast s}$ can be approximated as:
\begin{equation}
f_{\ast s}\cong \widehat{f_{s}}\left[ 1+h_{Ds}^{\left( 1\right)
}+h_{Ds}^{\left( 2\right) }\right] ,  \label{solo}
\end{equation}%
where the leading-order KDF $\widehat{f_{s}}$ does not depend on the
gradients of $\Lambda _{s}$. Hence, all the informations about the gradients
of the structure functions appear only through the first-order (in $%
\varepsilon $) perturbations $h_{Ds}^{\left( 1\right) }$ and $h_{Ds}^{\left(
2\right) }$. These are denoted respectively as the \emph{%
diamagnetic-correction} (see Ref.\cite{Catto1987}) and the \emph{%
energy-correction} (see Paper II), which result from the leading-order
Taylor expansions with respect to $\psi _{\ast s}$ and $\Phi _{\ast s}$. In
particular, the following results apply. First, $\widehat{f_{s}}$ is
expressed as%
\begin{eqnarray}
&&\left. \widehat{f_{s}}=\frac{n_{s}}{\left( 2\pi /M_{s}\right) ^{3/2}\left(
T_{\parallel s}\right) ^{1/2}T_{\perp s}}\right.  \notag \\
&&\times \exp \left\{ -\frac{M_{s}\left( \mathbf{v}-\mathbf{W}%
_{s}-U_{\parallel s}^{\prime }\mathbf{b}^{\prime }\right) ^{2}}{%
2T_{\parallel s}}-m_{s}^{\prime }\frac{B^{\prime }}{\Delta _{T_{s}}}\right\}
\label{solo3}
\end{eqnarray}%
and hence is identified with a \emph{bi-Maxwellian KDF with parallel
velocity perturbations }(see Paper II). In Eq.(\ref{solo3}) $\mathbf{W}%
_{s}=\Omega _{s}R^{2}\nabla \varphi $ and $U_{\parallel s}^{\prime }=\frac{%
I^{\prime }}{B^{\prime }}\xi _{s}$ are related to the leading-order toroidal
and parallel flow velocities and depend on angular frequencies of the
general form $\Omega _{s}=\Omega _{s}\left( \psi ,\Phi \right) $ and $\xi
_{s}=\xi _{s}\left( \psi ,\Phi \right) $. In addition, the function $n_{s}$
is defined in terms of the pseudo-density $\eta _{s}$ as%
\begin{equation}
n_{s}\left( \psi ,\Phi \right) \equiv \eta _{s}\left( \psi ,\vartheta ,\Phi
\right) \exp \left[ \frac{X_{s}}{T_{\parallel s}}\right]  \label{density}
\end{equation}%
and%
\begin{eqnarray}
X_{s} &\equiv &\left( M_{s}\frac{\left\vert \mathbf{W}_{s}\right\vert {}^{2}%
}{2}+\frac{Z_{s}e}{c}\psi \Omega _{s}-Z_{s}e\Phi +\Upsilon _{s}^{\prime
}\right) ,  \label{X} \\
\Upsilon _{s}^{\prime } &\equiv &\frac{M_{s}U_{\parallel s}^{\prime 2}}{2}%
\left( 1+\frac{2\Omega _{s}}{\xi _{s}}\right) +  \notag \\
&&+\left( \frac{M_{s}c\nabla ^{\prime }\psi ^{\prime }\cdot \nabla ^{\prime
}\Phi ^{\prime }}{B^{\prime 2}}+\frac{Z_{s}e}{c}\psi ^{\prime }\right) \xi
_{s}.  \label{y}
\end{eqnarray}%
Second, the diamagnetic and energy-correction contributions $h_{Ds}^{\left(
1\right) }$ and $h_{Ds}^{\left( 2\right) }$ are given by%
\begin{eqnarray}
h_{Ds}^{\left( 1\right) } &=&\left\{ \frac{cM_{s}R}{Z_{s}e}\left[ Y_{1}+Y_{3}%
\right] +\frac{M_{s}R}{T_{\parallel s}}\psi \Omega _{s}A_{3}\right\} \left(
\mathbf{v\cdot }\widehat{e}_{\varphi }\right) ,  \label{hd1} \\
h_{Ds}^{\left( 2\right) } &=&\frac{M_{s}}{2Z_{s}e}\left\{ Y_{4}-\frac{Z_{s}e%
}{T_{\parallel s}}\frac{\psi \Omega _{s}}{c}C_{3s}+\frac{p_{\varphi
s}^{\prime }\xi _{s}}{T_{\parallel s}}C_{5s}\right\} v^{2}.  \label{hd2}
\end{eqnarray}%
Here $Y_{i},$ $i=1,5$, is defined as
\begin{eqnarray}
Y_{1} &\equiv &\left[ A_{1s}+A_{2s}\left( \frac{H_{s}}{T_{\parallel s}}-%
\frac{1}{2}\right) -\mu _{s}^{\prime }A_{4s}\right] , \\
Y_{3} &\equiv &\left[ \frac{p_{\varphi s}^{\prime }\xi _{s}}{T_{\parallel s}}%
A_{5s}-A_{2s}\frac{p_{\varphi s}^{\prime }\xi _{s}}{T_{\parallel s}}\right] ,
\\
Y_{4} &\equiv &\left[ C_{1s}+C_{2s}\left( \frac{H_{s}}{T_{\parallel s}}-%
\frac{1}{2}\right) -\mu _{s}^{\prime }C_{4s}\right] ,
\end{eqnarray}%
where $H_{s}={E}_{s}-\frac{Z_{s}e}{c}\psi _{\ast s}\Omega _{s}$ and the
following definitions have been introduced: $A_{1s}\equiv \frac{\partial \ln
\beta _{s}}{\partial \psi },$ $A_{2s}\equiv \frac{\partial \ln T_{\parallel
s}}{\partial \psi },$ $A_{3s}\equiv \frac{\partial \ln \Omega _{s}}{\partial
\psi },$ $A_{4s}\equiv \frac{\partial \alpha _{s}}{\partial \psi },$ $%
A_{5s}\equiv \frac{\partial \ln \xi _{s}}{\partial \psi }$ and $C_{1s}\equiv
\frac{\partial \ln \beta _{s}}{\partial \Phi },$ $C_{2s}\equiv \frac{%
\partial \ln T_{\parallel s}}{\partial \Phi }$, $C_{3s}\equiv \frac{\partial
\ln \Omega _{s}}{\partial \Phi }$, $C_{4s}\equiv \frac{\partial \alpha _{s}}{%
\partial \Phi },$ $C_{5s}\equiv \frac{\partial \ln \xi _{s}}{\partial \Phi }$%
.

\bigskip

The outcome of the perturbative theory is as follows:

1) The asymptotic expansion in terms of $\psi _{\ast s}$ and leading to the
diamagnetic-correction $h_{Ds}^{\left( 1\right) }$ is formally analogous to
that presented in Ref.\cite{Catto1987}. The Taylor expansion in terms of $%
\Phi _{\ast s}$ (energy expansion) is instead a novel feature of the present
approach and leads to the energy-correction $h_{Ds}^{\left( 2\right) }$.

2) The kinetic equilibrium $f_{\ast s}$ is compatible with species-dependent
rotational frequencies $\Omega _{s}$ and $\xi _{s}$. No restriction follows
from the KDF on their relative magnitudes, so that the general ordering $%
\frac{\xi _{s}}{\Omega _{s}}\sim O\left( \varepsilon ^{0}\right) $ is
permitted.

3) A fundamental feature is related to the functional dependences imposed by
the kinetic constraints on the structure functions. As a basic consequence,
the latter depend both on the poloidal flux $\psi $ and the ES potential $%
\Phi $. As proved below, the ES potential $\Phi $ is generally a function of
the form $\Phi =\Phi (\mathbf{x},\varepsilon ^{k}t),$ with $\mathbf{x}%
=\left( R,z\right) $, i.e. it is not simply a $\psi $-flux function. Hence,
when expressed in magnetic coordinates, the structure functions become
generally of the form $\Lambda _{s}\equiv \overline{\Lambda _{s}}\left( \psi
,\vartheta ,\varepsilon ^{k}t\right) $. This type of functional dependence
is expected to apply for arbitrary nested magnetic surfaces having finite
inverse aspect ratio. On the other hand, in the case of large aspect ratio ($%
1/\delta \gg 1$), the poloidal dependences in $\Lambda _{s}$ are expected to
become negligible. Nevertheless, $h_{Ds}^{\left( 2\right) }$ remains finite
even in this case. The reason is that also in this limit the double Taylor
expansion (\ref{espan0}) still applies.

4) The coefficients $A_{is}$ and $C_{is}$, $i=1,5$, can be identified with
\textit{effective thermodynamic forces}, containing the spatial variations
of $\Lambda _{s}$ respectively across the $\psi =const.$ and $\Phi =const.$
surfaces.

\section{The Vlasov fluid approach}

An elementary consequence concerns the fluid approach defined in terms of
the Vlasov description, i.e., based on the moment equations following from
the asymptotic Vlasov kinetic equation (see Eq.(\ref{asvla})). In fact,
assuming that the KDF is identified with\textit{\ }the adiabatic invariant
given by Eq.(\ref{form}), these equations are \textit{necessarily all
identically satisfied} in an asymptotic sense, namely neglecting corrections
of $O\left( \varepsilon ^{n+1}\right) $. Furthermore, because $f_{\ast s}$
is by construction periodic, also the corresponding solubility conditions,
related to the requirement of periodicity in terms of the $\vartheta $%
-coordinate, are necessarily fulfilled. To prove these statements we notice
that if $Q(\mathbf{z})$ is an arbitrary weight function, identified for
example with $Q=\left( 1,\mathbf{v},v^{2}\right) ,$ then the generic moment
of Eq.(\ref{asvla}) is:%
\begin{equation}
\int_{\Gamma _{u}}d^{3}vQ\frac{d}{dt}f_{\ast s}=0+O\left( \varepsilon
^{n+1}\right) ,  \label{dd}
\end{equation}%
where $\Gamma _{u}$ denotes the appropriate velocity space of integration.
Using the chain rule, and taking into account explicitly also the dependence
in terms of $p_{\varphi s}^{\prime }$, this can be written as%
\begin{equation}
\int_{\Gamma _{u}}d^{3}vQ\left\{
\begin{array}{c}
\frac{d\psi _{\ast s}}{dt}\frac{\partial f_{\ast s}}{\partial \psi _{\ast s}}%
+\frac{dE_{s}}{dt}\frac{\partial f_{\ast s}}{\partial E_{s}}+ \\
+\frac{dm_{s}^{\prime }}{dt}\frac{\partial f_{\ast s}}{\partial
m_{s}^{\prime }}+\frac{dp_{\varphi s}^{\prime }}{dt}\frac{\partial f_{\ast s}%
}{\partial p_{\varphi s}^{\prime }}%
\end{array}%
\right\} =0+O\left( \varepsilon ^{n+1}\right) .
\end{equation}%
On the other hand, Eq.(\ref{dd}) can also be represented as%
\begin{equation}
\int_{\Gamma _{u}}d^{3}v\left\{ \frac{d}{dt}\left[ Qf_{\ast s}\right]
-f_{\ast s}\frac{d}{dt}Q\right\} =0+O\left( \varepsilon ^{n+1}\right) ,
\end{equation}%
which recovers the usual form of the velocity-moment equations in terms of
suitable (and \textit{uniquely defined}) fluid fields. For $Q=\left( 1,%
\mathbf{v}\right) $ one obtains, in particular, that the species continuity
and linear momentum fluid equations are satisfied identically up to
infinitesimals of $O\left( \varepsilon ^{n+1}\right) $. Similarly, the law
of conservation of the species total canonical momentum can be recovered by
setting $Q=\psi _{\ast s}$, namely%
\begin{equation}
\int_{\Gamma _{u}}d^{3}v\frac{d}{dt}\left[ \psi _{\ast s}f_{\ast s}\right]
=0+O\left( \varepsilon ^{n+1}\right) .
\end{equation}%
In the stationary case this implies the customary species angular momentum
conservation law for the species angular momentum $L_{s}^{tot}\equiv
M_{s}R^{2}n_{s}^{tot}\mathbf{V}_{s}^{tot}\cdot \nabla \varphi $ (see Paper
II). Here the notation is standard. In particular the velocity moments of
the KDF $\left\{ n_{s}^{tot},\mathbf{V}_{s}^{tot},\underline{\underline{\Pi }%
}_{s}^{tot},L_{cs}^{tot}\right\} $ can be introduced, to be referred to as
\textit{species number density, flow velocity, tensor pressure} and \textit{%
canonical toroidal momentum}. They are defined by the integrals $%
\int_{\Gamma _{u}}d^{3}vQf_{\ast s}$, where $Q$ is now identified
respectively with $Q=\left\{ 1,\frac{\mathbf{v}}{n_{s}^{tot}},M_{s}\left(
\mathbf{v}-\mathbf{V}_{s}^{tot}\right) \left( \mathbf{v}-\mathbf{V}%
_{s}^{tot}\right) ,\frac{Z_{s}e}{c}\psi _{\ast s}\right\} $. It is worth
remarking here that \textit{the velocity moments are unique once the KDF} $%
f_{\ast s}$ [see Eq.(\ref{sol2})] \textit{is prescribed in terms of the
structure functions} $\left\{ \Lambda _{\ast s}\right\} .$ On the other
hand, as a result of Eqs.(\ref{asvla}) and (\ref{dd}), it follows that the
stationary fluid moments calculated in terms of the KDF $f_{\ast s}$ are
identically solutions of the corresponding stationary fluid moment equations.

We conclude this section pointing out that no restrictions can possibly be
required on the KDF and the EM potentials as a consequence of the validity
of these moment equations. Therefore, \textit{the only possible constraints
on the KDF are necessarily only those arising from the solubility conditions
of the Maxwell equations.}

\section{Constitutive equations for species number density and flow velocity}

In this section we present the leading-order expressions of the species
number density and flow velocities predicted by the kinetic equilibrium. The
calculation of these fluid moments is required for the subsequent analysis
of the Maxwell equations. An explicit calculation of the moment integrals
can be carried out by adopting the perturbative asymptotic expansion of $%
f_{\ast s}$ described in Section 6. This also requires to perform an \textit{%
inverse GK transformation}, by expressing all of the guiding-center
quantities appearing in the equilibrium KDF in terms of the actual particle
position, according to Eq.(\ref{trGK}).

Consider first the evaluation of the species flow velocity $\mathbf{V}%
_{s}^{tot}$. Adopting the GK representation for the particle velocity, the
leading-order contribution to the flow velocity is found to be%
\begin{eqnarray}
\mathbf{V}_{s} &\cong &\mathbf{U+}\frac{I}{B}\left( \Omega _{s}+\xi
_{s}\right) \mathbf{b}+  \notag \\
&&+\frac{T_{\perp s}}{T_{\parallel s}}\left[ R^{2}\left( \Omega _{s}+\xi
_{s}\right) \nabla \varphi -\mathbf{U}\right] \cdot \left( \underline{%
\underline{\mathbf{1}}}-\mathbf{bb}\right) ,
\end{eqnarray}%
where $\mathbf{U}$ is the frozen-in velocity defined by Eq.(\ref{u}). Then,
ignoring correction of $O\left( \varepsilon \right) $, $\mathbf{U}$ can be
approximated as $\mathbf{U}\cong R^{2}\Omega _{o}\mathbf{\nabla }\varphi
\cdot \left( \underline{\underline{\mathbf{1}}}-\mathbf{bb}\right) .$ Here $%
\Omega _{o}$ is the \textit{species-independent} and $\psi $\textit{-flux
function} \cite{Catto1987}%
\begin{equation}
\Omega _{0}\left( \psi ,\varepsilon ^{k}t\right) \equiv c\frac{\partial
\left\langle \Phi \right\rangle }{\partial \psi }  \label{omega0}
\end{equation}%
and $\left\langle \Phi \right\rangle =\varkappa ^{-1}\oint \frac{d\vartheta
}{\mathbf{B}\cdot \nabla \vartheta }\Phi $ denotes the $\psi $-surface
average, with $\varkappa ^{-1}\equiv \oint \frac{d\vartheta }{\mathbf{B}%
\cdot \nabla \vartheta }$. Then, in terms of the relative toroidal frequency
$\Delta \Omega _{s}\equiv \Omega _{s}-\Omega _{o},$ the leading-order flow
velocity becomes%
\begin{eqnarray}
\mathbf{V}_{s} &\cong &\left[ \Omega _{o}+\frac{T_{\perp s}}{T_{\parallel s}}%
\left[ \Delta \Omega _{s}+\xi _{s}\right] \right] R^{2}\nabla \varphi +
\notag \\
&&+\left[ \Delta \Omega _{s}+\xi _{s}\right] \frac{I}{B}\left( 1-\frac{%
T_{\perp s}}{T_{\parallel s}}\right) \mathbf{b}.  \label{Vstot1}
\end{eqnarray}%
This implies that $\mathbf{V}_{s}$ can be decomposed in terms of the \emph{%
total toroidal and poloidal rotation velocities}
\begin{eqnarray}
V_{Ts}\left( \psi ,\vartheta ,\Phi \right) &\equiv &R\Omega _{Ts}=\mathbf{V}%
_{s}\cdot \mathbf{e}_{\varphi },  \label{tor flow} \\
V_{Ps}\left( \psi ,\vartheta ,\Phi \right) &\equiv &\frac{\Omega _{Ps}}{%
\left\vert \nabla \vartheta \right\vert }=\mathbf{V}_{s}\cdot \mathbf{e}_{P},
\label{pol flow}
\end{eqnarray}%
where $\mathbf{e}_{P}\equiv \frac{\nabla \vartheta }{\left\vert \nabla
\vartheta \right\vert }$, and the corresponding rotation frequencies $\Omega
_{Ts}$ and $\Omega _{Ps}$ are respectively:%
\begin{eqnarray}
\Omega _{Ts} &=&\Omega _{o}+\frac{T_{\perp s}}{T_{\parallel s}}\left[ \Delta
\Omega _{s}+\xi _{s}\right] +  \notag \\
&&+\left[ \Delta \Omega _{s}+\xi _{s}\right] \frac{I^{2}}{B^{2}R^{2}}\left(
1-\frac{T_{\perp s}}{T_{\parallel s}}\right) , \\
\Omega _{Ps} &=&\left[ \Delta \Omega _{s}+\xi _{s}\right] \frac{I}{B^{2}J}%
\left( 1-\frac{T_{\perp s}}{T_{\parallel s}}\right) ,
\end{eqnarray}%
with $\frac{1}{J}\equiv \nabla \psi \times \nabla \varphi \cdot \nabla
\vartheta $.

We remark that:

1) To leading-order in $\varepsilon $, the poloidal flow velocity (\ref{pol
flow}) is non-zero \emph{only in the presence of temperature anisotropy. }%
More precisely, provided $\frac{T_{\perp s}}{T_{\parallel s}}\neq 1,$ a
non-vanishing $V_{Ps}$ may arise only if $\Delta \Omega _{s}+\xi _{s}\neq 0.$
Therefore, even if $\Omega _{s}$ coincides with the frozen-in frequency $%
\Omega _{0}$, $\Omega _{Ps}$ is different from zero if $\xi _{s}\neq 0$.

2) The effect of the contributions $\Delta \Omega _{s}$ and $\xi _{s}$ is
analogous, although their physical origins are different. In particular $%
\Delta \Omega _{s}$ represents the departure from the frozen-in rotation
velocity $\Omega _{o}$, while $\xi _{s}$ determines the parallel velocity
perturbation in the KDF.

3) If the frozen-in condition is invoked, namely $\Omega _{s}\equiv \Omega
_{o}$, Eq.(\ref{Vstot1}) becomes%
\begin{equation}
\mathbf{V}_{s}\cong \Omega _{o}R^{2}\nabla \varphi +\xi _{s}\left[ \frac{%
T_{\perp s}}{T_{\parallel s}}R^{2}\nabla \varphi +\frac{I}{B}\left( 1-\frac{%
T_{\perp s}}{T_{\parallel s}}\right) \mathbf{b}\right] ,  \label{Correction1}
\end{equation}%
which takes into account both finite poloidal rotation and temperature
anisotropy. In case of isotropic temperatures, i.e., $\frac{T_{\perp s}}{%
T_{\parallel s}}=1,$ the previous equation provides a purely toroidal flow
given by $\mathbf{V}_{s}\cong \left( \Omega _{o}+\xi _{s}\right) R^{2}\nabla
\varphi .$ When $\xi _{s}\equiv 0$ this reduces to the customary result \cite%
{Catto1987}, namely $\mathbf{V}_{s}\cong \Omega _{o}R^{2}\nabla \varphi $.

Finally, we report the calculation of the number density $n_{s}^{tot}$.
Neglecting again first-order diamagnetic and energy-correction
contributions, the leading-order species number density is found to be%
\begin{equation}
n_{s}=\eta _{s}\left( \psi ,\Phi \right) \exp \left[ \frac{\widehat{X}%
_{s}-Z_{s}e\Phi }{T_{\parallel s}}\right] ,  \label{density2}
\end{equation}%
where%
\begin{eqnarray}
\widehat{X}_{s} &\equiv &\frac{M_{s}}{2}\frac{I^{2}}{B^{2}}\left( \Omega
_{s}+\xi _{s}\right) ^{2}-\frac{M_{s}}{2}U^{2}+  \notag \\
&&+\left[ M_{s}R^{2}\mathbf{U\cdot \nabla }\varphi \mathbf{+}\frac{Z_{s}e}{c}%
\psi \right] \Omega _{s}+  \notag \\
&&+\left[ \frac{M_{s}}{B}\frac{c\nabla \psi \cdot \nabla \Phi }{B}\mathbf{+}%
\frac{Z_{s}e}{c}\psi \right] \xi _{s}+  \notag \\
&&+\frac{M_{s}}{2}\frac{T_{\perp s}}{T_{\parallel s}}\left( \Delta \Omega
_{s}+\xi _{s}\right) ^{2}\left[ R^{2}-\frac{I^{2}}{B^{2}}\right]  \label{X^}
\end{eqnarray}%
contains the combined contribution of the kinetic energies carried by the
rotation frequencies $\Omega _{s},$ $\xi _{s}$ and the frozen-in velocity $%
\mathbf{U}$.

\section{Quasi-neutrality}

In this section we investigate the implications of the quasi-neutrality
condition following from the Poisson equation. Here by quasi-neutrality we
mean that the equation%
\begin{equation}
\sum\limits_{s}Z_{s}en_{s}^{tot}=0  \label{qneu}
\end{equation}%
is satisfied asymptotically in the sense that%
\begin{equation}
\frac{\left\vert \nabla \cdot \mathbf{E}\right\vert }{\left\vert
\sum\limits_{s}Z_{s}en_{s}^{tot}\right\vert }\sim \frac{O\left( \varepsilon
_{D}^{2}\right) }{O\left( \varepsilon \right) },
\end{equation}%
with $\varepsilon _{D}\equiv \frac{\lambda _{D}}{\Delta L}\ll 1$ denoting
the Debye-length dimensionless parameter, with $\lambda _{D}\sim \lambda
_{Ds}=\tau _{ps}v_{ths}$, $\Delta L\sim \Delta L_{s}=\Delta t_{s}v_{ths}$,
and $n_{s}^{tot}$ the total species-number density. We intend to show that
the first two terms in the Laurent expansion (\ref{Laurent}) of $\Phi $ can
be determined from Eq.(\ref{qneu}) by prescribing $n_{s}^{tot}$ to
leading-order in $\varepsilon $, namely in terms of Eq.(\ref{density2}). In
particular the following result holds.

\textbf{THM.1 - Explicit form of the ES potential }$\Phi $\textbf{.}

\textit{Let us assume that the species KDF is defined by Eq.(\ref{sol2}) and
the finite aspect-ratio ordering applies. Then, imposing the
quasi-neutrality condition (\ref{qneu}) in the case of a two-species
ion-electron plasma, the following propositions hold:}

T1$_{1}$) \textit{Correct through }$O\left( \varepsilon ^{0}\right) $,
\textit{the ES potential satisfies the asymptotic implicit equation}%
\begin{equation}
\Phi \simeq \frac{S\left( \psi ,\vartheta ,\Phi \right) }{e\left( \frac{Z_{i}%
}{T_{\parallel i}}+\frac{1}{T_{\parallel e}}\right) },  \label{Fi}
\end{equation}%
\textit{where }$S\left( \psi ,\vartheta ,\Phi \right) $\textit{\ is the
source term given by}%
\begin{equation}
S\left( \psi ,\vartheta ,\Phi \right) \equiv \ln \left( \frac{\eta _{e}}{%
Z_{i}\eta _{i}}\right) +\left[ \frac{\widehat{X}_{e}}{T_{\parallel e}}-\frac{%
\widehat{X}_{i}}{T_{\parallel i}}\right] ,
\end{equation}%
\textit{with }$\eta _{s}$\textit{\ being the species pseudo-density and the
quantity }$\widehat{X}_{s}=\widehat{X}_{s}\left( \psi ,\vartheta ,\Phi
\right) $\textit{\ defined by Eq.(\ref{X^}).}

T1$_{2}$) \textit{If the temperatures are non-isotropic, then the species
pseudo-density is generally of the form }$\eta _{s}=\eta _{s}\left( \psi
,\vartheta ,\Phi \right) $. \textit{Instead, in the case of isotropic
temperatures, }$\eta _{s}=\eta _{s}\left( \psi ,\Phi \right) $\textit{.}

T1$_{3}$) \textit{A particular solution consistent with the kinetic
constraints is obtained letting }$Z_{i}\eta _{i}=\eta _{e}$.

T1$_{4}$) \textit{In particular, in validity of T1}$_{3}$\textit{, correct
through }$O\left( \varepsilon ^{0}\right) $ \textit{the ES potential }$\Phi $%
\textit{\ is uniquely determined by Eq.(\ref{Fi}) and is necessarily of the
form (\ref{Laurent}),} \textit{where\ }$\Phi _{-1}$\textit{\ obeys the
equation}%
\begin{equation}
\Phi _{-1}\left( \psi \right) \cong \frac{\psi \left[ \frac{Z_{i}\left(
\Omega _{i}+\xi _{i}\right) }{T_{\parallel i}}+\frac{\Omega _{e}+\xi _{e}}{%
T_{\parallel e}}\right] }{c\left( \frac{Z_{i}}{T_{\parallel i}}+\frac{1}{%
T_{\parallel e}}\right) },  \label{fi-1}
\end{equation}%
\textit{while }$\Phi _{0}$\textit{\ is obtained subtracting }$\Phi _{-1}$%
\textit{\ from Eq.(\ref{Fi}).}

PROOF - T1$_{1}$ - The proof of the first statement can be obtained from Eq.(%
\ref{qneu}) by substituting for the species number density the leading-order
solution given by Eq.(\ref{density2}). T1$_{2}$ - The proof follows by
noting that, in validity of the kinetic constraint on $\beta _{\ast s}$, the
species pseudo-density is such that $\frac{\eta _{s}}{T_{\perp s}}=\frac{%
\eta _{s}}{T_{\perp s}}\left( \psi ,\Phi \right) $. On the other hand, from
the kinetic constraint imposed on $\alpha _{\ast s}$ and the prescriptions
that $B=B\left( \psi ,\vartheta \right) $ and $T_{\parallel s}=T_{\parallel
s}\left( \psi ,\Phi \right) $, it must be that $T_{\perp s}$ is necessarily
of the type $T_{\perp s}=T_{\perp s}\left( \psi ,\vartheta ,\Phi \right) $.
Therefore, the general dependence of the pseudo-density is also necessarily
of the form $\eta _{s}=\eta _{s}\left( \psi ,\vartheta ,\Phi \right) $. On
the other hand, in the limit of isotropic temperatures $T_{\perp
s}=T_{\parallel s}=T_{s}\left( \psi ,\Phi \right) $ and $\alpha _{\ast s}=0$%
. The functional dependence of $\eta _{s}$ becomes therefore of the type $%
\eta _{s}=\eta _{s}\left( \psi ,\Phi \right) $. T1$_{3}$ - Due to the
arbitrariness of the structure function $\beta _{\ast s}$, it follows that $%
\beta _{e}$ and $\beta _{i}$ can always be defined in such a way that$\frac{%
\beta _{e}}{\beta _{i}}\frac{T_{\perp e}}{T_{\perp i}}=1$ even when $%
T_{\perp i}\neq T_{\perp e}$. In particular, this constraint is consistent
with the requirement that the ES potential vanishes identically in the
absence of toroidal and poloidal rotations. T1$_{4}$ - By definition the
Poisson equation, subject to suitable boundary conditions, must determine
completely (i.e., uniquely) the ES potential $\Phi $. Therefore, Eq.(\ref{Fi}%
) yields necessarily the complete solution, correct through $O\left(
\varepsilon ^{0}\right) $. In particular, by inspecting the order of
magnitude of the different contributions in the source term $\widehat{X}%
_{s}\left( \psi ,\vartheta ,\Phi \right) $, the Laurent expansion (\ref%
{Laurent}) can be introduced. In particular, by retaining in $\widehat{X}%
_{s}\left( \psi ,\vartheta ,\Phi \right) $ only contributions of $1/O\left(
\varepsilon \right) $, $\Phi _{-1}\left( \psi \right) $ is found to obey Eq.(%
\ref{fi-1}). \textbf{Q.E.D.}

A fundamental implication of THM.1, and in particular of the validity of Eq.(%
\ref{Laurent}), is to assure the consistency of the perturbative $\sigma
_{s} $-expansion as well as the orderings introduced in Sections 3 and 4. In
fact, let us inspect the order of magnitude (with respect to the parameter $%
\varepsilon $) of the r.h.s. of Eq.(\ref{fi-1}). For definiteness, let us
assume that $\left( \Omega _{i}+\xi _{i}\right) \sim \left( \Omega _{e}+\xi
_{e}\right) \sim \Omega _{0}$, requiring $T_{\parallel i}/T_{\parallel
e}\sim O\left( \varepsilon ^{0}\right) $ and $Z_{i}\sim O\left( \varepsilon
^{0}\right) $. Due to Eq.(\ref{omega0}) it follows that the order of
magnitude of $\Phi _{-1}$ is $\Phi _{-1}\sim \psi \frac{\Omega _{0}}{c}$. On
the basis of this conclusion, the following statement holds.

\textbf{Corollary to THM.1 - Consistency with the }$\sigma _{s}$\textbf{%
-expansion.}

\textit{Given validity of THM.1 and the quasi-neutrality condition, invoking
the previous assumptions it follows that }$\sigma _{i}\sim \sigma _{e}\sim
O\left( \varepsilon \right) $\textit{.}

PROOF - First, by assumption $\sigma _{i}\sim \left\vert \frac{\frac{M_{i}}{2%
}v_{thi}^{2}}{{Z_{i}e}\Phi }\right\vert $ and $\sigma _{e}\sim \left\vert
\frac{\frac{M_{e}}{2}v_{the}^{2}}{{e}\Phi }\right\vert $. As a consequence,
due to the previous hypotheses $\sigma _{i}\sim \sigma _{e}$. Furthermore,
thanks to quasi-neutrality, it follows that%
\begin{equation}
\sigma _{i}\sim \left\vert \frac{M_{s}v_{thi}\frac{1}{2}v_{thi}}{{Z_{i}e}%
\psi \frac{\Omega _{0}}{c}}\right\vert \sim \left\vert \frac{M_{s}v_{thi}R}{%
\frac{{Z_{i}e}\psi }{c}}\right\vert \left\vert \frac{\frac{1}{2}v_{thi}}{%
\Omega _{0}R}\right\vert .
\end{equation}%
The order of magnitude of the two factors on the r.h.s follows from the
asymptotic ordering for the canonical-momentum parameter and the requirement
indicated above that $\Omega _{0}R\sim v_{thi}$ (see also Ref.\cite%
{Catto1987}). It is concluded that, since by construction $O\left(
\varepsilon _{i}\right) \sim O\left( \varepsilon \right) $, $\left\vert
\frac{M_{s}v_{thi}R}{\frac{{Z_{i}e}\psi }{c}}\right\vert \sim O\left(
\varepsilon \right) $, while $\left\vert \frac{\frac{1}{2}v_{thi}}{\Omega
_{0}R}\right\vert \sim O\left( \varepsilon ^{0}\right) $, which manifestly
implies the thesis. \textbf{Q.E.D.}

The following further remarks are useful in order to gain insight in the
previous results.

1) Eq.(\ref{Fi}) represents the general solution holding in the case of a
two-species plasma characterized by temperature anisotropy, poloidal and
toroidal flow velocities.

2) In Eq.(\ref{fi-1}) all quantities $\Lambda _{s}^{\left( 1\right) }\equiv
\left\{ \Omega _{s},\xi _{s},T_{\parallel s}\right\} $ can be considered (to
leading-order in $\varepsilon $) as only $\psi $-functions, namely of the
form $\Lambda _{s}^{\left( 1\right) }=\Lambda _{s}^{\left( 1\right) }\left(
\psi \right) $. Therefore, Eq.(\ref{fi-1}) provides an ODE for $\Phi
_{-1}\left( \psi \right) $.

\section{The Ampere equation}

Let us now investigate the constraints imposed by the Ampere law on the
leading-order current densities and equilibrium flows. Let us consider the
case of a two-species plasma. The following results apply.

\textbf{THM.2 - Constraints on poloidal and toroidal flows.}

\textit{Given validity of the asymptotic Vlasov kinetic equation (\ref{asvla}%
) for the species KDF defined by Eq.(\ref{sol2}), the quasi-neutrality
condition (\ref{qneu}) and the magnetized-plasma asymptotic orderings (see
Section 3), for a two-species plasma the following propositions hold:}

T2$_{1}$) \textit{The poloidal flow velocity }$V_{Ps}\left( \psi ,\vartheta
,\Phi \right) $ \textit{may be either species-dependent or independent. In
the first case necessarily the constraint condition}%
\begin{equation}
\frac{\partial }{\partial \vartheta }\left[ \sum\limits_{s}Z_{s}en_{s}V_{Ps}%
\left( \psi ,\vartheta ,\Phi \right) \right] =0  \label{solub}
\end{equation}%
\textit{\ must be fulfilled. In the second case, if Eq.(\ref{solub}) is not
satisfied, the corresponding total equilibrium current density must vanish
identically.}

T2$_{2}$) \textit{In both cases, the toroidal flow velocity remains
species-dependent, so that the corresponding current density is generally
non-vanishing.}

T2$_{3}$) \textit{Both poloidal and toroidal magnetic fields can be
self-generated by the plasma.}

PROOF - T2$_{1}$ - Let us consider first the component of Ampere's equation
along the directions orthogonal to $\nabla \varphi $. This yields the
following set of two scalar equations for the toroidal magnetic field $%
I\nabla \varphi $:%
\begin{eqnarray}
\frac{\partial I}{\partial \psi } &=&\frac{4\pi }{c}\sum%
\limits_{s}Z_{s}en_{s}V_{Ps}\left( \psi ,\vartheta ,\Phi \right) \left[
1+O\left( \varepsilon \right) \right] , \\
\frac{\partial I}{\partial \vartheta } &=&0+O\left( \varepsilon \right) ,
\end{eqnarray}%
implying manifestly the solubility condition (\ref{solub}). Therefore,
either the total poloidal current density is a $\psi $-function, or the
poloidal flow velocity $V_{Ps}\left( \psi ,\vartheta ,\Phi \right) $ must be
species-independent. The first condition can always be satisfied by suitably
selecting the species pseudo-density. In fact, even in validity of T1$_{3}$,
the species pseudo-density can be defined in such a way to satisfy the
constraint (\ref{solub}). Therefore, excluding the null solution, a
non-vanishing current density must appear when $V_{Ps}$ is species-dependent.

T2$_{2}$ - The proof of the second statement follows by noting that, in
validity of proposition T2$_{1}$, the quantity $\frac{T_{\perp s}}{%
T_{\parallel s}}\left[ \Delta \Omega _{s}+\xi _{s}\right] $ may still remain
species-dependent. As a consequence, by direct inspection of Eq.(\ref{tor
flow}), it follows that the toroidal current density is generally non-null.

T2$_{2}$ - Thanks to the previous propositions, it follows that both the
toroidal and poloidal magnetic fields can be self generated. In particular,
the self toroidal field requires necessarily the presence of temperature
anisotropy, while the poloidal self field may arise even in the case of
isotropic temperature, due to deviations from the frozen-in condition $%
\Omega _{s}=\Omega _{o}$ and/or parallel velocity perturbations associated
to $\xi _{s}$. \textbf{Q.E.D.}

We briefly mention the case of a multi-species plasma. In fact, in
collisionless systems plasma sub-species can be introduced, simply based on
the topology of their phase-space trajectories. For example, different
species can be identified distinguishing between circulating and
magnetically-trapped particles. These components can in principle be
characterized by KDFs carrying different structure functions, and in
particular different poloidal flow velocities. In this case both the
poloidal and toroidal flow velocities remain generally species-dependent.
Therefore, the corresponding current densities may be expected to be
non-vanishing.

\bigskip

\section{Comparisons with literature}

An interesting issue is related to comparisons with the literature. For what
concerns the kinetic formulation, the relevant benchmark is represented by
Ref.\cite{Catto1987}, where the theory of collisional transport in
toroidally rotating plasmas was investigated. Although the conceptual
foundations of the perturbative kinetic approach here adopted have already
been exhaustively detailed in Sections 2 to 10, it is worth analyzing some
differences arising between the two approaches. In detail, besides the
inclusion of temperature anisotropy, parallel and toroidal velocity
perturbations as well as the prescription of the kinetic constraints, the
main differences with Ref.\cite{Catto1987} are as follows:

1) The first one lies in the choice of equilibrium KDF. This is related, in
particular, to the different definition adopted here for the dynamical
variable $H_{\ast s}$ (see Eq.(\ref{H*})). The motivation for this
definition have been detailed in Section 5. Such a choice permits to obtain
an explicit analytical solution for the leading-order ES potential $\Phi
_{-1}\left( \psi \right) $ (see THM.1), based uniquely on the
quasi-neutrality condition rather than imposing fluid constraints (see Ref.%
\cite{Catto1987}).

2) In the present approach \emph{no constraints arising from the moment
(i.e., fluid) equations} are placed on the structure functions $\left\{
\Lambda _{\ast s}\right\} $ [see Eq.(\ref{kincon1})] and consequently on the
velocity moments of the KDF $f_{\ast s}$. In particular, in our case, unlike
the case of collisional plasmas treated in Ref.\cite{Catto1987}, the general
form of the equilibrium species-fluid velocity $\mathbf{V}_{s}$ is merely a
consequence of the form prescribed for the equilibrium KDF. Therefore, it
cannot follow from imposing the validity of fluid equations, but only from
the solubility conditions of the Maxwell equations.

3) The analysis of the Ampere equation has been carried out to investigate
its consequences on the toroidal and poloidal species-flow velocities in the
presence of temperature anisotropy (see THM.2). The discussion extends the
treatment given in Ref.\cite{Catto1987}, where only differential toroidal
flows were retained in the kinetic treatment.

Let us now consider, for the sake of reference, also the case of statistical
fluid approaches. Such treatments (including those adopting multi-fluid
formulations) typically do not\emph{\ }rely on kinetic closures conditions
and/or include FLR as well as perturbative kinetic effects, such as
diamagnetic and energy-correction contributions. Further issues include:

1) The treatment of kinetic constraints. As shown here, kinetic constraints
are critical for the construction of the KDF. They allow the structure
functions to retain, in principle, both $\psi $ (leading-order) and $%
\vartheta $ (first-order) dependences. The correct functional form of the
fluid fields, arising as a consequence of the kinetic constraints, may not
be correctly retained in customary fluid treatments (see for example Refs.%
\cite{Mc2001,Mc2011}).

2) The proper inclusion of slowly time-dependent temperature and pressure
anisotropies. As pointed out here, the functional form of the parallel and
perpendicular temperature is related to microscopic conservation laws, in
particular particle magnetic moment conservation. On the other hand, fluid
approaches normally ignore such constraints. Even when kinetic closure
conditions are invoked for the pressure tensor (see for example Ref.\cite%
{Mc2011}), their validity may become questionable if they are not based on
consistent equilibrium solutions for the KDF.

3) Another example-case is provided by the kinetic prescription for the
expression of the number density, here shown to exhibit a complex dependence
in terms of the ES potential, centrifugal potential as well as toroidal and
parallel frequencies (for comparison see Ref.\cite{Mc2011}).

4) Finally, the functional form of the poloidal flow velocity may differ
from what can be obtained adopting a two-fluid approach \cite{Mc2001}. In
particular, in our treatment the toroidal and parallel rotation frequencies
are considered independent of each other, so that kinetic constraints need
to be imposed separately on $\Omega _{s}$ and $\xi _{s}$. Furthermore,
according to the kinetic treatment, a non-vanishing equilibrium poloidal
flow velocity can only appear in the presence of temperature anisotropy.

\bigskip

\section{Concluding remarks}

In this paper, a theoretical formulation of quasi-stationary configurations
for collisionless and axisymmetric Tokamak plasmas has been presented. This
is based on a kinetic approach developed within the framework of the
Vlasov-Maxwell description. It has been shown that a new type of asymptotic
kinetic equilibria exists, which can be described in terms of generalized
bi-Maxwellian distributions. By construction, these are expressed in terms
of the relevant particle first integrals and adiabatic invariants. Such
solutions permit the consistent treatment of a number of physical properties
characteristic of collisionless plasmas. They include, in particular,
differential toroidal rotation and finite temperature anisotropy and
poloidal flows in non-uniform multi-species Tokamak plasmas subject to
intense quasi-stationary magnetic and electric fields. The existence of
these solutions has been shown to be warranted by imposing appropriate
kinetic constraints for the structure functions which appear in the species
distribution functions. By construction, the theory assures the validity of
the fluid moment equations associated to the Vlasov equation. In particular,
the novelty of the approach lies in the explicit construction of asymptotic
solutions for the fluid equations in terms of constitutive equations for the
fluid fields. The approach is based on a perturbative asymptotic expansion
of the equilibrium distribution function, which allows also the
determination of diamagnetic and energy-correction contributions. The latter
are found to be linearly proportional to suitable effective thermodynamic
forces. Finally, the constraints placed by the Maxwell equations have been
investigated. As a result, the electrostatic potential has been determined
by imposing the quasi-neutrality condition. Furthermore, it has been shown
that non-trivial solutions for the toroidal and poloidal species rotation
frequencies are allowed consistent with the solubility conditions arising
from the Ampere law. The discussion presented here provides a useful
background for future investigations.

\begin{acknowledgments}
This work has been partly developed in the framework of MIUR (Italian
Ministry of University and Research) PRIN Research Programs and the
Consortium for Magnetofluid Dynamics, Trieste, Italy.
\end{acknowledgments}


\bigskip

\begin{thebibliography}{99}
\bibitem{Eriksson1997} L.G. Eriksson, E. Righi and K.-D. Zastrow, Plasma
Phys. Controlled Fusion \textbf{39}, 27 (1997).

\bibitem{Zhou2010} D. Zhou, Phys. Plasmas \textbf{17}, 102505 (2010).

\bibitem{Boedo2011} J.A. Boedo, E.A. Belli, E. Hollmann, W.M. Solomon, D.L.
Rudakov, J.G. Watkins, R. Prater, J. Candy, R.J. Groebner, K.H. Burrell,
J.S. deGrassie, C.J. Lasnier, A.W. Leonard, R.A. Moyer, G.D. Porter, N.H.
Brooks, S. Muller, G. Tynan and E.A. Unterberg, Phys. Plasmas \textbf{18},
032510 (2011).

\bibitem{Catto2009} G. Kagan and P.J. Catto, Phys. Plasmas \textbf{16},
056105 (2009).

\bibitem{Pamela2010} S. Pamela, G. Huysmans and S. Benkadda, Plasma Phys.
Controlled Fusion \textbf{52}, 075006 (2010).

\bibitem{Hameiri1983} E. Hameiri, Phys. Fluids \textbf{26}, 230 (1983).

\bibitem{Hassam96} A.B. Hassam, Nucl. Fusion \textbf{36}, 707 (1996).

\bibitem{Max1998} L.-J. Zheng and M. Tessarotto, Phys. Plasmas \textbf{5},
1403 (1998).

\bibitem{Brizard1994} F.L. Hinton, J. Kim, Y.-B. Kim, A. Brizard and K.H.
Burrel, Phys. Rev. Lett. \textbf{72}, 1216 (1994).

\bibitem{Catto1987} P.J. Catto, I.B. Bernstein and M. Tessarotto, Phys.
Fluids B \textbf{30}, 2784 (1987).

\bibitem{Max1992} M. Tessarotto and R.B. White, Phys. Fluids B: Plasma
Physics \textbf{4}, 859 (1992).

\bibitem{Max1996} M. Tessarotto, J.L. Johnson, R.B. White and L.-J. Zheng,
Phys. Plasmas \textbf{3}, 2653 (1996).

\bibitem{Max1997} M. Tessarotto, M. Pozzo, L.-J. Zheng and R. Zorat, Rivista
del Nuovo Cimento \textbf{20}, 1 (1997).

\bibitem{Briz1995} A. Brizard, Phys. Plasmas \textbf{2}, 459 (1995).

\bibitem{Cr2010} C. Cremaschini, J.C. Miller and M. Tessarotto, Phys.
Plasmas \textbf{17}, 072902 (2010).

\bibitem{Cr2011} C. Cremaschini, J.C. Miller and M. Tessarotto, Phys.
Plasmas \textbf{18}, 062901 (2011).

\bibitem{Bern1985} I.B. Bernstein and P.J. Catto, Phys. Fluids \textbf{28},
1342 (1985).

\bibitem{Little1979} R.G. Littlejohn, J. Math. Phys. \textbf{20}, 2445
(1979).

\bibitem{Little1981} R.G. Littlejohn, Phys. Fluids\textbf{\ 24}, 1730 (1981).

\bibitem{Kruskal} M. Kruskal, J. Math. Phys. Sci. \textbf{3}, 806 (1962).

\bibitem{Mc2001} K.G. McClements and A. Thyagaraja, Mon. Not. R. Astron.
Soc. \textbf{323}, 733-742 (2001).

\bibitem{Mc2011} M.J. Hole, G. Von Nessi, M. Fitzgerald, K.G. McClements, J.
Svensson and the MAST team, Plasma Phys. Control. Fusion \textbf{53}, 074021
(2011).
\end{thebibliography}
\end{document}